\documentclass[twocolumn,showpacs,aps,prl,superscriptaddress]{revtex4}
\usepackage{graphicx}
\usepackage{dcolumn}
\usepackage{epsfig}
\usepackage{amsmath}

\newcommand{\BaBarYear}    {06}
\newcommand{\BaBarNumber}  {028}
\newcommand{\SLACPubNumber} {11834}

 \newcommand{\BaBarType}      {PUB}  % Journal publication

\input babarsym   % the official way
\RequirePackage{xspace}

\newcommand{\pvec}{{\bf p}}
\newcommand{\calB}{\ensuremath{{\cal B}}}

\newcommand{\bfemsix}{${\cal B}(10^{-6}$)}

\newcommand{\DE}{\ensuremath{\Delta E}}

\newcommand{\xf}{\ensuremath{{\cal F}}}

\newcommand{\thetaT}{\ensuremath{\theta_{\rm T}}}
\newcommand{\costhr}{\ensuremath{\cos\thetaT}}

\newcommand{\dEdx}{\ensuremath{\mathrm{d}E/\mathrm{d}x}}

\newcommand\etal{{\it et al.}}
\newcommand{\half}{\ensuremath{{1\over2}}}

\newcommand{\bma}[1]{\boldmath{$#1$}}

\newcommand{\bfig}{\begin{figure}[htbpc!]}
\newcommand{\efig}{\end{figure}}
\newcommand\bef{\begin{figure}}
\newcommand\edf{\end{figure}}
\newcommand\dbline{\noalign{\vskip 0.10truecm\hrule}\noalign{\vskip 2pt}\noalign{\hrule\vskip 0.10truecm}}
\providecommand{\tbline}{\noalign{\vskip 0.05truecm\hrule\vskip0.05truecm}}

\newcommand\beq{\begin{equation}}
\newcommand\eeq{\end{equation}}
\newcommand\bear{\begin{array}}
\newcommand\enar{\end{array}}
\newcommand\beqa{\begin{eqnarray}}
\newcommand\eeqa{\end{eqnarray}}
\newcommand\ben{\begin{enumerate}}
\newcommand\een{\end{enumerate}}

\newcommand{\UfourS}{\ensuremath{\Upsilon(4S)}}

\newcommand{\etapepp}{\ensuremath{\etapr_{\eta\pi\pi}}}

\newcommand{\etaprg}{\ensuremath{\etapr_{\rho\gamma}}}

   \newcommand{\rhoz}{\ensuremath{\rho^0}}

\newcommand{\fetapreppetapreppks}{\ensuremath{\etapr_{\eta \pi\pi} \etapr_{\eta \pi\pi} \KS}}
\newcommand{\fetapreppetapreppkz}{\ensuremath{\etapr_{\eta \pi\pi} \etapr_{\eta \pi\pi} K^0}}
\newcommand{\fetapreppetaprrgks}{\ensuremath{\etapr_{\eta \pi\pi} \etapr_{\rho \gamma} \KS}}
\newcommand{\fetapreppetaprrgkz}{\ensuremath{\etapr_{\eta \pi\pi} \etapr_{\rho \gamma} K^0}}
\newcommand{\fetapreppetaprrgkp}{\ensuremath{\etapr_{\eta \pi\pi}\etapr_{\rho\gamma} K^{+}}}
\newcommand{\fetapreppetapreppkp}{\ensuremath{\etapr_{\eta \pi\pi}\etapr_{\eta\pi\pi} K^{+}}}

\newcommand{\fetapretaprkz}{\ensuremath{\etapr \etapr K^0}}

\newcommand{\fetapretaprk}{\ensuremath{\etapr \etapr K}}
\newcommand{\fetapretaprkp}{\ensuremath{\etapr \etapr K^+}}

\newcommand{\psiKs}{\mbox{$B^0\ra J/\psi  K^0_S $}}
\newcommand{\KsKsKs}{\mbox{$B^0\ra K^0_S K^0_S  K^0_S $}}
\providecommand{\BetapKzs}{\mbox{$B^0 \rightarrow \eta^{\prime} \KS $}}

\newcommand{\signf}{$\cal S$ ($\sigma$)}
\newcommand{\eff}{$\epsilon$ (\%)}

\begin{document}

%**% \preprint{\babar-PUB-\BaBarYear/\BaBarNumber} 
%**% \preprint{SLAC-PUB-\SLACPubNumber} 

\begin{flushleft}
\babar-\BaBarType-\BaBarYear/\BaBarNumber \\
SLAC-PUB-\SLACPubNumber \\
%**% hep-ex/\LANLNumber
\end{flushleft}

\title{
 \large \bf\boldmath Search for $B$ Meson Decays to \fetapretaprk 
}

%% author list as of 01-Mar-2006 (610 authors)
%
\author{B.~Aubert}
\author{R.~Barate}
\author{M.~Bona}
\author{D.~Boutigny}
\author{F.~Couderc}
\author{Y.~Karyotakis}
\author{J.~P.~Lees}
\author{V.~Poireau}
\author{V.~Tisserand}
\author{A.~Zghiche}
\affiliation{Laboratoire de Physique des Particules, F-74941 Annecy-le-Vieux, France }
% ...
\collaboration{The \babar\ Collaboration}
\noaffiliation

\date{\today}

\begin{abstract}
We describe searches for  decays of $B$ mesons to
the charmless final states  \fetapretaprk. The data consist of
228 million  \BB\ pairs produced in 
\epem\ annihilation, collected with the
\babar\ detector at the Stanford Linear Accelerator Center. The  
90\%  confidence level upper limits for the branching fractions are
$\calB(\Bz \to \fetapretaprkz) <31 \times 10^{-6} $ and 
 $\calB (\Bp \to \fetapretaprkp) <25 \times  10^{-6} $.
\end{abstract}


\maketitle

The phenomenon of \CP\ violation  has been  extensively studied in recent 
years at the $B$ factories. The observations of mixing-induced \CP\ violation
in \psiKs\ decays  \cite{MixInd} and of direct \CP\ violation both in the neutral
kaon system \cite{kaon} and in  $B^0\ra K^+\pi^-$
decays \cite{dir} are in  agreement with expectations in the Standard Model 
(SM)  of electroweak interactions~\cite{SM}. 
Some possible evidence of 
disagreement between experimental results and  SM expectations is found in 
$B$ decay modes dominated by penguin amplitudes, for example in 
 the decay \BetapKzs~\cite{peng}. Further 
important information  about  \CP violation and hadronic $B$ decays
can be provided by 
the measurements of branching fractions and time-dependent \CP\ asymmetries 
in $B$ decays to three-body final states containing two identical neutral 
particles of spin zero and another spin zero neutral particle~\cite{Tim}. An
example of such a decay 
is \KsKsKs, which  has already been observed~\cite{KsKsKs}. 
Since  the branching fractions for the decays $B\ra \eta^{\prime} K$ are
large~\cite{peng},
another example which might be particularly interesting for time-dependent 
\CP\ violation analysis is the mode $\Bz \to \fetapretaprkz$. 

We present the results of searches for  the exclusive decay modes
 $\Bp \to \fetapretaprkp$~\cite{ChargeCon}  and  \Bz \to \fetapretaprkz,
 which are studied for the first time.
The results are based on data collected
with the \babar\ detector~\cite{BABARNIM}
at the PEP-II asymmetric-energy $e^+e^-$ collider~\cite{pep}
located at the Stanford Linear Accelerator Center. The analyses use
an integrated luminosity of 207~\invfb, corresponding to 
$228$ million \BB\ pairs, recorded at the
$\Upsilon (4S)$ resonance (center-of-mass energy $\sqrt{s}=10.58\
\gev$).

Charged particles from the \epem\ interactions are detected, and their
momenta measured, by a combination of five layers of double-sided
silicon microstrip detectors and a 40-layer drift chamber,
both operating in the 1.5~-T magnetic field of a superconducting
solenoid. Photons and electrons are identified with a CsI(Tl) crystal
electromagnetic calorimeter (EMC).  Further charged particle
identification (PID) is provided by the average energy loss (\dEdx) in
the tracking devices and by an internally reflecting, ring-imaging
Cherenkov detector (DIRC) covering the central region.
A $K/\pi$ separation of better than four standard deviations ($\sigma$)
is achieved for momenta below 3~\gevc, decreasing to 2.5~$\sigma$ at the
highest momenta in the $B$ decay final states.

\begin{table*}[t]
\caption{
Fitted signal yield, fit bias, detection
efficiency  \eff, daughter branching fraction product $\prod\calB_i$,
significance \signf\  , measured branching
fraction \calB\ with statistical error for each decay mode. For the
combined measurements we give
the significance  (with systematic uncertainties included) and the  branching fraction
with statistical and systematic uncertainty (in parentheses the  90\%
CL upper limit).
}
\label{tab:results}
\begin{tabular}{lccccccc}
\dbline
Mode& \quad Yield \quad&\quad Fit bias (ev) &\quad \eff \quad &\quad
$\prod\calB_i$ (\%) \quad&\quad \signf \quad &\quad \bfemsix \quad  \\
\tbline
~~\fetapreppetapreppkz\ &   $0.9^{+1.4}_{-0.7}$&$+0.5$ &$2.9$&$1.1$&$0.5$&$6^{+20}_{-10}$\\
~~\fetapreppetaprrgkz\  &   $4.1^{+8.1}_{-6.7}$ &$+3.8$&$3.7$&$3.6$&$0.0$&$1^{+27}_{-22}$\\
\bma{\fetapretaprkz\ }&  &  &  & &\bma{0.5 } & \bma{5^{+14}_{-9} \pm 1 \quad(<31) }   \\
\tbline
~~\fetapreppetapreppkp\ &   $4.2^{+3.7}_{-2.8}$  &$+0.5$&$3.5$&$3.1$&$2.1$&$15^{+15}_{-11}$ \\ 
~~\fetapreppetaprrgkp\ &   $13.6^{+11.7}_{-10.1}$  &$+8.5$&$3.4$&$10.4$&$0.5$&$6^{+15}_{-13}$ \\
\bma{\fetapretaprkp\ } &  &  &  & &\bma{2.0} & \bma{11^{+9}_{-7} \pm 1 \quad(<25)  } \\
\dbline
\end{tabular}
\vspace{-5mm}
\end{table*}

The $B$ daughter candidates are reconstructed through their decays
$\etapr\ra\eta\pip\pim$ (\etapepp), where
$\eta\ra\gaga$, and  
$\etapr\ra\rhoz\gamma$ (\etaprg), where $\rhoz\ra\pip\pim$. We require
the laboratory energy of the photons to be  greater than 
30~\mev\ for \etapepp and  200~\mev\ for \etaprg.
We impose the following requirements on the invariant mass (in \mevcc) of
the candidate final states:
$490 < m(\gamma\gamma) < 600$ for $\eta$,
$930 < m(\pip\pim\eta) <990$ for \etapepp,
$930 < m(\pip\pim\gamma) <990$ for \etaprg, and   $510 < m(\pip\pim)
<1000$ for \rhoz.
%Table \ref{tab:rescuts}\ lists the
%requirements on the invariant mass of these particles' final states.
Secondary tracks in \etapr\ candidates are rejected if
 their PID signatures from the DIRC and \dedx\ are consistent with those for 
protons, kaons, or electrons. Charged $K$ candidates are selected if their 
PID signatures from the DIRC and \dedx\ are consistent with that for kaons. 
Candidate \KS\ decays are formed from pairs of oppositely charged tracks 
with $486 < m(\pip\pim)<510$ \mevcc,  a decay vertex
$\chi^2$ probability larger than $0.001$,  and a reconstructed decay length 
greater than three times  its uncertainty.  

We reconstruct the $B$ meson candidate by combining two  
\etapr\ candidates and a charged or neutral kaon. We consider
only cases with two \etapepp\ candidates or a  \etapepp\ and a
\etaprg. We do not 
consider the case with two \etaprg\ candidates due to the high
background present in this mode. 
From the kinematics of the \UfourS\ decays we determine the energy-substituted mass
$\mes=\sqrt{\frac{1}{4}s-\pvec_B^2}$
and the energy difference $\DE = E_B-\half\sqrt{s}$, where
$(E_B,\pvec_B)$ is the $B$ meson 4-momentum vector, and
all values are expressed in the \UfourS\ frame.
The resolution is $3.0\ \mevcc$ for \mes\ and $26\ \mev$ for \DE.  We
require $5.25\ \gevcc<\mes<5.29\ \gevcc$ and $|\DE|<0.2$ GeV.

Backgrounds arise primarily from random combinations of particles in
continuum $\epem\ra\qqbar$ events ($q=u,d,s,c$).  We reduce these with
requirements on the angle
\thetaT\ between the thrust axis of the $B$ candidate in the \UfourS\
frame and that of the rest of the charged tracks and neutral calorimeter
clusters in the event.  The distribution is sharply
peaked near $|\costhr|=1$ for \qqbar\ jet pairs,
and nearly uniform for $B$ meson decays.  The requirement 
is $|\costhr|<0.9$ \,($|\costhr|<0.7$ for the charged mode with \etaprg).
We define the decay angle $\theta^{\rho}_{\rm dec}$ for the $\rho$ 
meson as the
angle between the momenta of a daughter particle and the
$\eta^{\prime}$, measured in the $\rho$  meson  rest frame. We require
for the \etaprg\  decay  $|\cos\theta^{\rho}_{\rm dec}|< 0.9$.
Events are retained only if they contain one or more charged tracks
that are not used in the candidate decay.

We obtain the signal event yields from unbinned extended maximum likelihood
fits.  The  input observables are \DE, \mes, the invariant masses of the two
\etapepp, a Fisher
discriminant \xf~\cite{Fisher}, and the 
variable  $|\cos\theta^{\rho}_{\rm dec}|$.
The Fisher discriminant \xf\  combines four variables: the angles, with
respect to the  
beam axis, of the $B$ momentum and the $B$ thrust axis 
(in the \UfourS\ frame), and the zeroth and second angular moments $L_{0,2}$ 
of the energy flow about the $B$ thrust axis.  The moments are defined by
$ L_j = \sum_i p_i\times\left|\cos\theta_i\right|^j$,
where $\theta_i$ is the angle, with respect to the $B$ thrust axis, of 
track or neutral cluster $i$, and $p_i$ is its momentum. The sum
excludes the $B$ candidate daughters.
  
The average number of candidates found per 
selected event is in the range 1.5 to 1.8, depending on the final
state.  We choose the candidate with the highest $B$ vertex  $\chi^2$
probability. From simulated events we find that
this algorithm selects the correct candidate in about
82\% of the events containing multiple candidates, and 
introduces negligible bias.

We use Monte Carlo (MC) simulation to estimate backgrounds from other $B$ decays, including 
final states with and without charm. These contributions are negligible for 
the \etapepp\ modes. For \etaprg\ we include a \BB component in the fit. 
We consider four categories in the likelihood fit: signal, 
self-cross feed (SCF) signal, defined as a signal candidate where one 
$B$ candidate daughter  has been exchanged with a particle from the rest of the event, and
 continuum    and \BB\ backgrounds. 

For each event $i$ and category  $j$,
the likelihood function is
\begin{equation}
L=  e^{-\left(\sum n_j\right)} \prod_{i=1}^N \left[\sum_{j=1}^4 n_j  {\cal P}_j ({\bf x}_i)\right] ,
\end{equation}
where  $N$ is the number of candidates,  $n_j$ is the number of events in 
category $j$, and     ${\cal P}_j ({\bf x}_i)$ is the corresponding
probability density function (PDF), evaluated   
with the observables ${\bf x}_i$ of the $i$th event. 
Since   correlations among the observables are small,
we take each  ${\cal P}$  as the product of the PDFs for the separate
variables. 
We determine the PDF parameters from Monte Carlo simulation
\cite{geant} of the 
signal, SCF,  and \BB\ background, while using 
  \mes\  and \DE\ sideband data ($5.25 < \mes\
<5.27$ \gevcc, $0.1<|\DE |<0.2$ \gev )  
to model the PDFs of  continuum  background. 

We parameterize each of the functions ${\cal P}(\mes)$, 
${\cal  P}(\DE)$, ${\cal P}( m_{\etapr })$, and 
$ {\cal P}( m_{\eta})$  for signal and  SCF with two Gaussian
distributions.  
The \mes\ distribution for  \BB\ and  continuum background is  described by a
threshold  function~\cite{argus}. 
The \DE\ distribution for \BB\ and  continuum background and  
 the $|\cos\theta^{\rho}_{\rm dec}|$ distributions are represented by linear or 
quadratic functions.
The distributions of $ m_{\etapr }$ and
$m_{\eta}$ in \BB\ and continuum background are described
by a Gaussian plus  linear function.
 The distribution of \xf\  is described with 
an asymmetric Gaussian function with a 
different width below and above the peak.
We allow  the continuum background PDF
parameters  to vary in the fit.
Large control samples of  $B\ra D(K\pi\pi)\pi$ decays 
are used to verify the simulated  \DE\ and \mes\ resolution. 

In Table~\ref{tab:results} we show the fitted signal yield, the fit bias in events, the detection
efficiency, the product of daughter branching fractions for each decay mode,
the significance \signf, and the measured branching fraction. 
We compute the branching fractions from the fitted  signal event
yields, detection efficiencies, 
daughter branching fractions, and number of produced $B$ mesons,
assuming equal production  
rates of charged and neutral $B$ meson pairs. We correct the yield for a
fit bias estimated with the simulations. 
We combine results from different sub-decay modes by adding the values
of $-2\ln{L}$ ,  taking proper  account of
 the correlated and uncorrelated systematic uncertainties.
We report the statistical significance and 
branching fraction for the individual decay channels. For 
the combined measurements we also report the 90\% confidence level (CL) 
upper limit. 
The statistical error on the signal yield is  the change in 
the central value when the quantity $-2\ln{L}$ increases by one 
unit from its minimum value. The significance is  the square root 
of the difference between the value of $-2\ln{L}$ (with systematic 
uncertainties included) for zero signal and the value at its minimum.
The 90\%  CL upper limit is taken to be the branching 
fraction below which lies 90\% of the total  likelihood integral 
in the positive branching fraction region.

\begin{figure}[!h]
\includegraphics[scale=0.40]{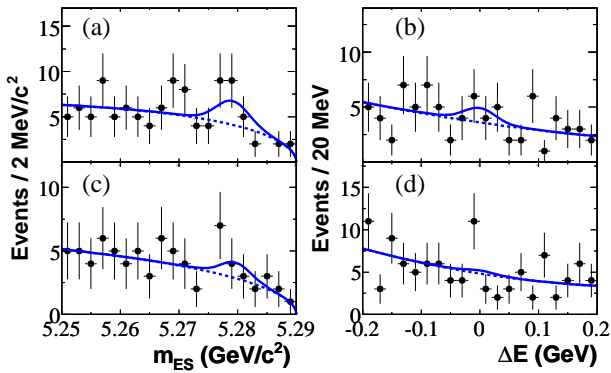}
\caption{
The $B$ candidate \mes\ and \DE\ projections for $\etapr \etapr \Kp$
(a, b) and $\etapr \etapr \KS\ $ (c, d) for the combined sub-decay
modes. Points with errors represent the data, solid curves the full fit 
functions and dashed curves the background functions. These plots are 
made with a requirement on the likelihood and thus do not show all
events in the data samples.
}
\label{f:projections}
\end{figure}

Figure~\ref{f:projections} shows projections of charged and neutral  
$\etapr\etapr K$ candidates onto \mes\ and \DE\ for the 
subset of candidates for which the signal likelihood
(computed without the variable plotted) exceeds a mode-dependent
threshold that optimizes the sensitivity.

\begin{figure}[!h]
\resizebox{\columnwidth}{!}{
\begin{tabular}{cc}
\includegraphics[]{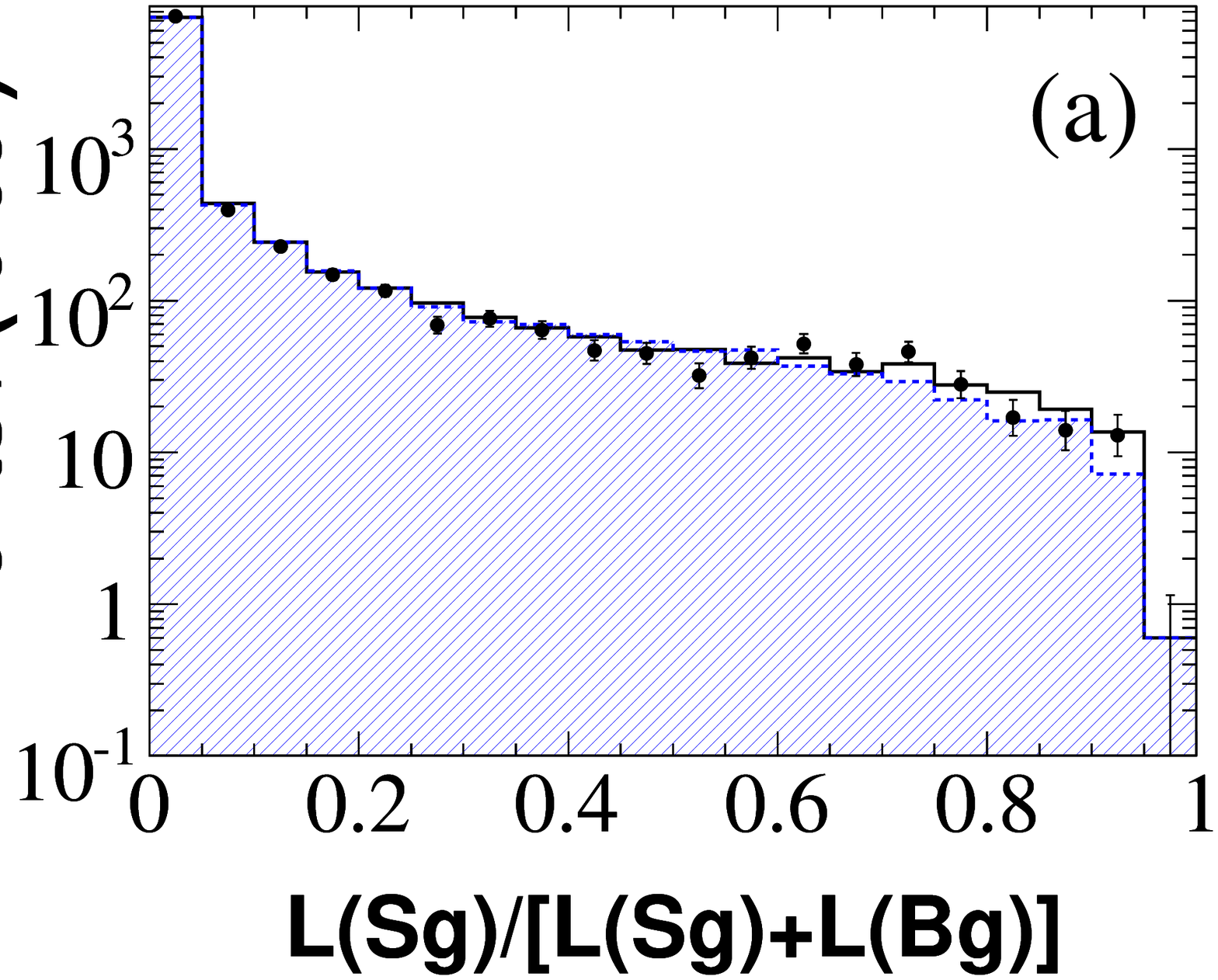} & \hspace{1cm}
\includegraphics[]{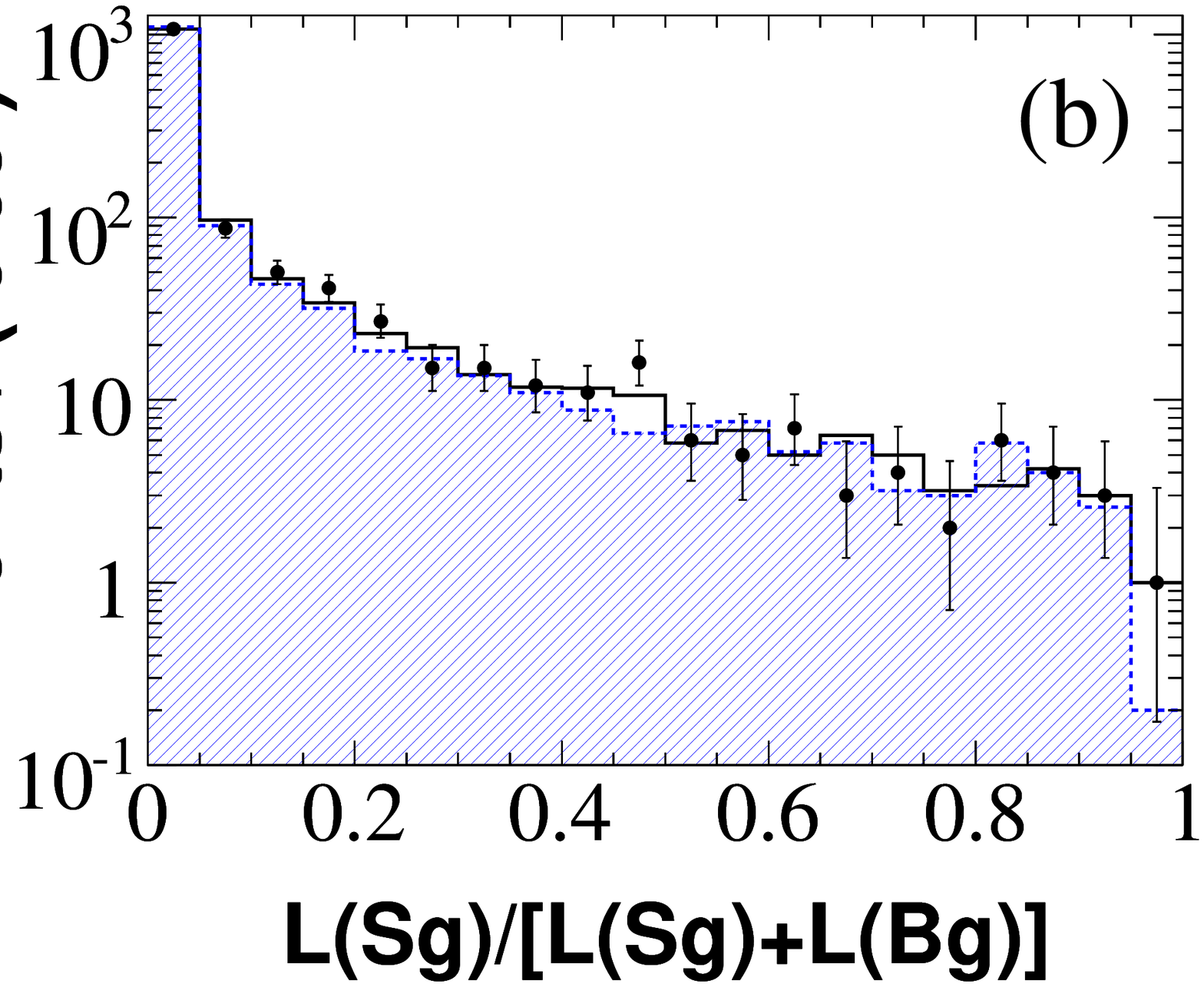} \\
\vspace{1cm}\\
\includegraphics[]{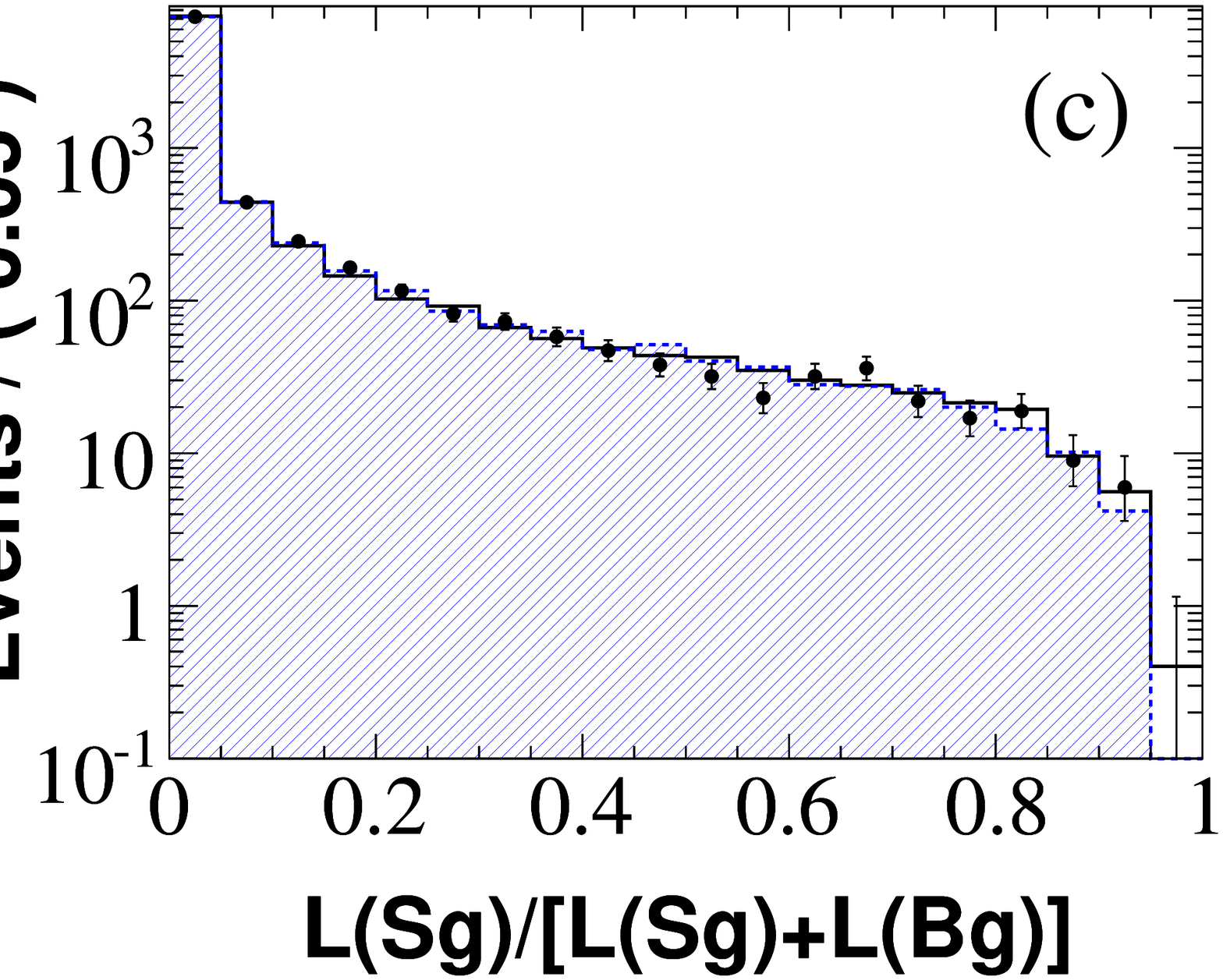}& \hspace{1cm}
\includegraphics[]{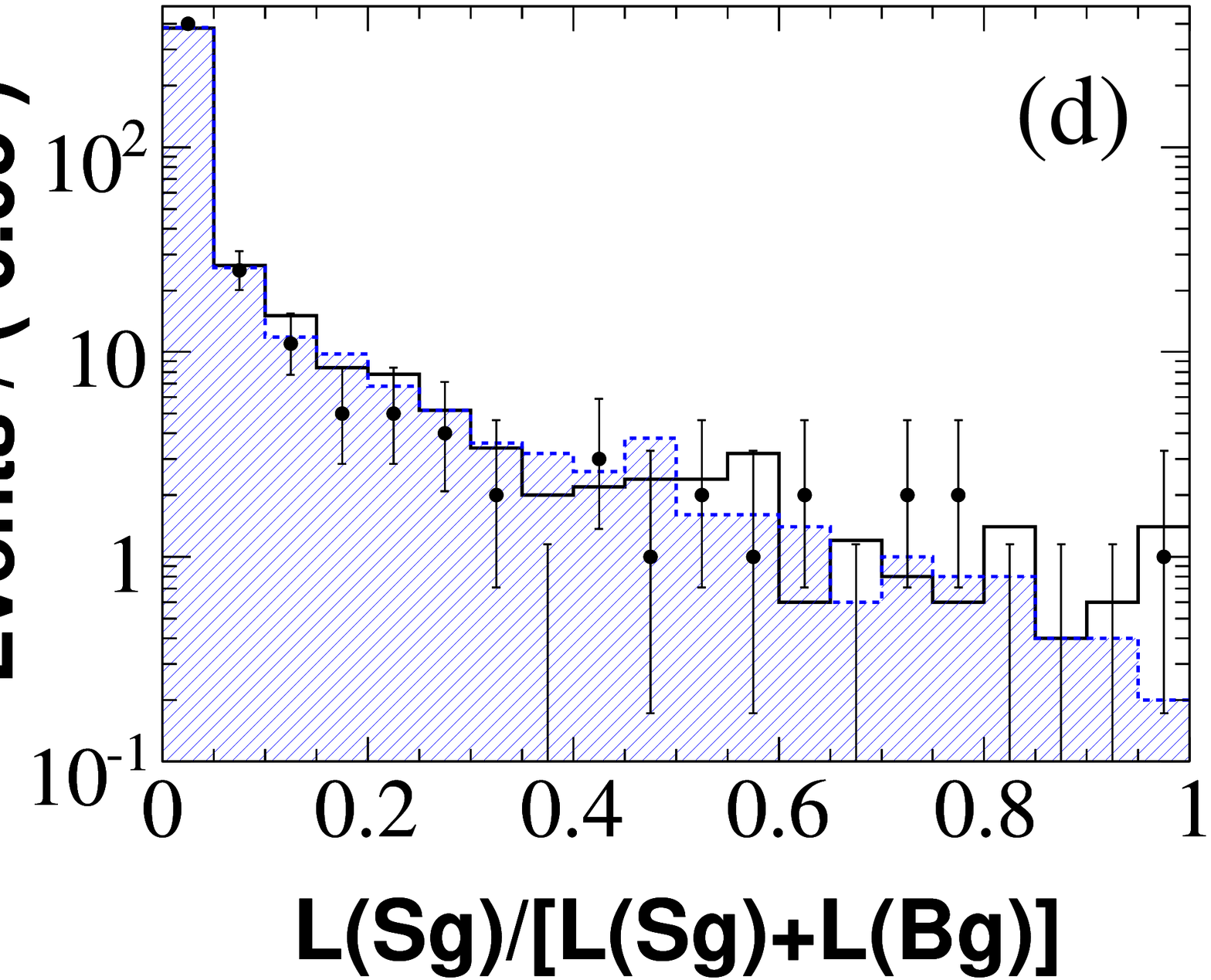} \\
\end{tabular}
}
\caption{
  The likelihood ratio L(Sg)/[L(Sg)+ L(Bg)] for
  the sub-decay modes of $\etapr \etapr K$: (a) \fetapreppetaprrgkp,
  (b) \fetapreppetapreppkp, (c) \fetapreppetaprrgks, 
  (d) \fetapreppetapreppks.
  The on-resonance data are shown as points with error bars;
  the sum of all simulated background samples is shown by the 
  shaded (dashed-line) histograms; and the sum of these backgrounds 
  plus the signal from the PDF model are given by the open  
  (solid-line) histograms.
}

\label{fig:LKRatio}
\end{figure}

The goodness-of-fit is further demonstrated by the distribution of the
likelihood ratio between the likelihood L(Sg) for the signal category  and the
sum of the likelihoods  for signal and all background categories
L(Bg) for data
and for simulation generated from the PDF model, shown 
in Figure~\ref{fig:LKRatio}. We
see good agreement between the model and the data. The
background is concentrated near zero, while any signal would appear
in a peak near one. 

The main sources of systematic errors include uncertainties in the
 PDF  parameters  and the maximum likelihood fit bias.    For the
signal, the uncertainties in the  PDF parameters are estimated by comparing MC and data in control samples.  Varying the
signal PDF parameters within these uncertainties, we estimate yield
uncertainties  up to 1 event, depending on the mode.  The uncertainty
from the fit bias is taken as half the correction
itself (up to 4 events). 
Uncertainties in our knowledge of the efficiency, found from auxiliary
studies, include $0.8\%\times N_t$ and $1.5\%\times N_\gamma$, where
$N_t$ and $N_\gamma$ are the numbers of tracks and photons, respectively,
in the $B$ candidate. A systematic uncertainty of 1.8\% is assigned to
 single photon  reconstruction efficiency.
There is a  systematic error of 2.1\% in the efficiency 
of \KS\ reconstruction and  3.0\% per $\eta$ in the efficiency of $\eta$
reconstruction.
  The uncertainty in the total number of \BB\ pairs in the
data sample is 1.1\%.  Published data~\cite{PDG2004}\ provide the
uncertainties in the $B$ daughter product branching fractions (3.5-4.9\%).

In conclusion, we have measured 90\% CL  upper 
limits for the branching fractions: 
$\calB(\Bz \to \fetapretaprkz) <31 \times 10^{-6}$ and 
 $\calB (\Bp \to \fetapretaprkp) <25  \times 10^{-6}$. From these results
we conclude that no \CP\ study is feasible in these $B$ decays with the currently
 available data samples.

We are grateful for the excellent luminosity and machine conditions
provided by our \pep2\ colleagues, 
and for the substantial dedicated effort from
the computing organizations that support \babar.
The collaborating institutions wish to thank 
SLAC for its support and kind hospitality. 
This work is supported by
DOE
and NSF (USA),
NSERC (Canada),
IHEP (China),
CEA and
CNRS-IN2P3
(France),
BMBF and DFG
(Germany),
INFN (Italy),
FOM (The Netherlands),
NFR (Norway),
MIST (Russia), and
PPARC (United Kingdom). 
Individuals have received support from CONACyT (Mexico), 
Marie Curie EIF (European Union),
the A.~P.~Sloan Foundation, 
the Research Corporation,
and the Alexander von Humboldt Foundation.

% Bibliography
%

\renewcommand{\baselinestretch}{1}


\begin{thebibliography}{99}
\bibitem{ChargeCon}
Charge-conjugate modes are implied throughout.
\end{thebibliography}
\end{document}